\documentclass[aps,preprint]{revtex4}

\def\be{\begin{eqnarray}}
\def\ee{\end{eqnarray}}
\def\nn{\nonumber}

\def\M{{\cal M}}
\def\l{\lambda}
\def\d{\delta}
\def\r2{\sqrt{2}}
\def\rt{\sqrt{3}}
\def\rs{\sqrt{6}}
\def\rtw{\sqrt{12}}
\def\G{{\cal G}}
\def\h{{1\over 2}}
\def\hh{{\rt\over 2}}
\def\({\left(}
\def\){\right)}

\begin{document}
\title{Mass Independent Textures and Symmetry}
\author{C.S. Lam}
\address{Department of Physics, McGill University,\\ Montreal, QC, Canada H3A 2T8\\
and\\
Department of Physics and Astronomy, University of British Columbia,  Vancouver, BC, Canada V6T 1Z1 \\
\email: Lam@physics.McGill.ca}

\begin{abstract}
A mass-independent texture is a set of linear relations of the fermion mass-matrix elements which 
imposes no constraint on the fermionic masses nor the Majorana phases. 
Magic and 2-3 symmetries are examples. 
We discuss the general construction and the properties
of these textures, as well as their relation to the quark and neutrino mixing matrices. 
Such a texture may be regarded as a symmetry, whose unitary generators of the
symmetry group can be explicitly constructed. In particular, the
symmetries connected with the tri-bimaximal neutrino mixing matrix are discussed, together
with the physical consequence of breaking one symmetry but preserving another.

\end{abstract}
\maketitle

\section{Introduction}
It was found by Harrison, Perkins, and Scott \cite{HPS} 
that neutrino mixing could be described by the tri-bimaximal
PMNS matrix
\be U_{HPS}=\pmatrix{{2 \over \rs}&{1\over\sqrt{3}}&0\cr -{1\over\sqrt{6}}&{1\over\sqrt{3}}&{1\over \sqrt{2}}\cr 
-{1\over\sqrt{6}}&{1\over\sqrt{3}}&-{1\over\sqrt{2}}\cr}\ee
whose mixing angles are given by $\sin^2\theta_{13}=0$, $\sin^2\theta_{12}=0.333$, and
$\sin^2\theta_{23}=0.50$. These values agree
very well with the numbers 
$\sin^2\theta_{13}=0.9{+2.3 \atop -0.9}\times 10^{-2}$, $\sin^2\theta_{12}=0.314(1 {+0.18\atop -0.15})$,
 and $\sin^2\theta_{23}=0.44(1 {+0.41\atop -0.22})$ obtained 
from a global fit of the experimental data \cite{FLMP}.

The symmetric Majorana neutrino mass matrix $M$, in the basis of a diagonal charged lepton matrix, is related to the 
PMNS unitary mixing
matrix  $U=\{u_{ij}\}$ and the complex neutrino masses $m={\rm diag}(m_1,m_2,m_3)$ by
\be
M=UmU^T=\{\sum_k u_{ik}m_ku_{jk}\}.\ee
The mass matrix $M_{HPS}$ obtained by taking $U=U_{HPS}$ obeys a 2-3 symmetry \cite{CSL1,O23} and a magic symmetry \cite{CSL2,MAGIC1, MAGIC2}. The former refers to the invariance of $M_{HPS}$ 
under a simultaneous permutation of the second and third columns, and the second and third rows.
The latter refers to the equality of the sum of each row and the sum of each column. Moreover, the correspondence is one to one \cite{CSL1,CSL2} in that
any symmetric mass matrix which is 2-3 and magic symmetric will lead to the mixing matrix $U_{HPS}$.

{\it Textures} are linear relations between mass-matrix elements. They
 have been studied extensively, especially for those with a fixed number of 
zeros  \cite{TEXTURE}. They relate
the matrix elements of the diagonalization matrix to the masses, and also
 the Majorana phases for Majorana neutrinos. 
We study in this paper {\it mass-independent textures}, the type
in which no constraints whatsoever are imposed
on the masses nor the Majorana phases, so that whatever restrictions derived from them occur
 only among the diagonalization-matrix elements.
For example, the 2-3 and magic symmetries mentioned above are both mass-independent textures.
In fact, the 2-3 texture for $M$ gives rise to a bimaximal mixing, and its magic texture gives
rise to a trimaximal mixing \cite{CSL1,O23,CSL2,MAGIC1,MAGIC2}, neither of which has anything to do with neutrino masses 
nor Majorana phases. In light of the importance of these two textures for neutrino mixing, we thought it
worthwhile to carry out a general study of the construction and the properties of the mass-independent textures.
Whether they are also important for quark mixing remains to be seen.
When we mention a texture in what follows, we automatically mean a mass-independent texture unless stated otherwise.

We shall show that a symmetry for a mass matrix leads to a mass-independent texture, 
and a mass-independent texture gives rise to a symmetry. 
Thus a study of these textures is a study of symmetry. It should be mentioned that the symmetry here refers to
a horizontal symmetry of the Standard Model with a left-handed Majorana mass, in which
the sole isodoublet Higgs transforms like a singlet under the horizontal symmetry of fermions. In the
more general case when additional Higgs are introduced which transforms non-trivially under the horizontal symmetry,
the mass-matrix elements are linear combinations of the appropriate horizontal Clebsch-Gordan coefficients, 
and the Higgs expectation values which break the horizontal symmetry. If the number of arbitrary
parameters are smaller than the number of independent mass-matrix elements, then one or several linear relations exists
between the mass-matrix elements and a texture is present. However these textures may nor may not be mass-independent,
so they may or may not correspond to symmetries in the present sense. An example of such a texture which is 
mass independent is given at the end of Sec.~6.

The construction and the general properties of mass-independent textures will be given in the next section.
The relation between textures and symmetries will be discussed in Sec.~3.
Given a texture, the allowed form of the diagonalization matrix will be discussed in Sec.~4. 
It is generally rather restrictive unless a mass degeneracy exists. 
As a consequence, it is impossible for both types of quarks or both types
of leptons to share a common texture, except in an approximate manner in which certain masses are regarded
as degenerate. These discussions can be found in Sec.~5.
 We will discuss in Sec.~6 various ways symmetries can be assigned to the lepton mass matrices
for the neutrino mixing matrix in the tri-bimaximal form, and the physical consequence of breaking one symmetry
while preserving another.
Finally, a summarizing conclusion is presented in Sec.~7.

\section{Mass-Independent Textures}
We consider two kinds of mass matrices $M$, symmetric and hermitian. The Majorana
neutrino mass matrix $M_\nu$ is symmetric, but the charged fermion matrices $M_f=\M_f\M_f^\dagger\ (f=u,d,e)$  are
 hermitian, where $\M_f$ is the Dirac mass matrix for fermion $f$. 

A hermitian mass matrix $M$ can be diagonalized by a unitary matrix $V$, so that $V^\dagger MV=\Lambda
={\rm diag}(\l_1,\l_2,\l_3)$,
with $\l_i$ being the square masses.
If $v_i$ is the $i$th column of $V$, then the diagonalization relation is equivalent to an eigenvalue problem
\be Mv_i=\l_iv_i,\quad (i=1,2,3).\ee

Given a normalized column vector $w=(w_1,w_2,w_3)^T$, a mass-independent texture for $M$ can be constructed
by equating $w$ to one of its eigenvectors $v_i$. If $Mw=\l_iw$, then it follows that
\be
{1\over w_1}\sum_{k=1}^3M_{1k}w_k={1\over w_2}\sum_{k=1}^3M_{2k}w_k={1\over w_3}\sum_{k=1}^3M_{3k}w_k.\ee
These two relations define a texture. All three expressions in (4) are equal to $\l_i$, but these two relations
do not care what $\l_i$ is, so the texture relations are mass independent.

If $M$ is a symmetric mass matrix, then we can find a
unitary $V$ to make $V^TMV=\Lambda={\rm diag}(\l_1,\l_2,\l_3)$ diagonal. The $\l_i$'s here are generally complex: their norms
are the neutrino masses, and their phases are the Majorana phases, though one of them is unphysical.

If $v_i$ is the $i$th column vector of $V$, instead of (3) we now have $Mv_i=\l_iv_i^*$. 
We shall refer to $v_i$ as a {\it pseudo-eigenvector}. Instead of (4) we have
\be
{1\over w_1^*}\sum_{k=1}^3M_{1k}w_k={1\over w_2^*}\sum_{k=1}^3M_{2k}w_k={1\over w_3^*}\sum_{k=1}^3M_{3k}w_k.\ee
For real $w$'s, there is no difference between (4) and (5).

Let us consider two illustrative examples. 
If $w=(1,1,1)^T/\rt$, then (4) becomes $\sum_kM_{1k}=\sum_kM_{2k}=\sum_kM_{3k}$,
so this texture is just the magic symmetry when $M$ is a symmetric matrix. 
If $w=(0,1,-1)^T/\r2$, then (4)
requires $M_{12}=M_{13}$ and $M_{22}+M_{32}=M_{23}+M_{33}$. If $M$ is symmetric, then the last equality becomes
$M_{22}=M_{33}$, so this texture is just the 2-3 symmetry.

Suppose $w$ and $w'$ are two mutually orthogonal normalized column vectors. The texture imposed by asking both
$w$ and $w'$ to be eigenvectors of $M$ define a {\it full texture}, so named because in that case the third normalized
eigenvector $w''$ is also known. Up to a phase it is just the unique vector orthogonal to $w$ and $w'$.
The diagonalization matrix $V$ is then fully determined up to umimportant phases, and order of the eigenvectors,
provided there
is no mass degeneracy in $M$. The 2-3 and magic textures taken together define a full texture,
whose diagonalization matrix is just $V=U_{HPS}$ \cite{CSL2,MAGIC1}. We shall
sometimes refer to what is defined in (4) or (5) 
as a {\it simple texture} to distinguish it from a full texture.

It is clear from (4) that unless one of the $w_i$'s is zero, the only diagonal matrix with any
texture is a multiple of the identity matrix. If a $w_i$ is zero, 
$M_{ii}$ may be arbitary, but the other matrix elements of the diagonal matrix must be equal. The only texture 
enjoyed by a general diagonal matrix is the ones defined with two vanishing $w_i$'s. This characterization of
diagonal matrices with a non-trivial texture will be used later.

If $M_a$ and $M_b$ are hermitean mass matrices with the same texture, {\it i.e.,} both have $w$ as an eigenvector, then
their inverses, their product, and their linear combination all have $w$ as an eigenvector, so they
all have the same texture. We will refer to this property of textures as {\it closure}. 
If $w$ is real, then $M^*$ also has the same texture, and the symmetric mass matrices are closed as well.

\section{Texture and Symmetry}
Mass-independent textures may be regarded as symmetries, and vice versa. 
A unitary transformation $f\to G_ff$ of the left-handed charged
fermion $f\ (=u,d,e)$ leads to the tranformation $M_f\to G_f^\dagger M_fG_f$ of its hermitean mass matrix. Thus
$M_f$ is invariant and $G_f$  a symmetry if and only if
$M_fG_f=G_fM_f$. This calls for simultaneous eigenvectors of $G_f$ and $M_f$. Similarly, 
a unitary transformation $\nu\to G_\nu\nu$ of the left-handed Majorana neutrino
 leads to the tranformation $M_\nu\to G_\nu^T M_\nu G_\nu$ of its symmetric mass matrix. Thus
$M_\nu$ is invariant and $G_\nu$ is a symmetry if and only if
$M_\nu G_\nu=G_\nu^* M_\nu$. As a result, if $v_i$ is a pseudo-eigenvector of $M_\nu$, then
$G_\nu v_i$ is also a pseudo-eigenvector of $M_\nu$, making it possible for the pseudo-eigenvectors of $M_\nu$
to be the eigenvectors of $G_\nu$.

Given a symmetry $G$ of a mass matrix $M$, all the eigenvectors of $G$ are the (pseudo-) eigenvectors of $M$ if
$G$ has no degenerate eigenvalues. In that case $M$ possesses a full texture defined by the eigenvectors of $G$.
However, if $G$ is doubly degenerate, then only its non-degenerate
eigenvector defines a (simple) texture for $M$. If it is triply degenerate, then it is a multiple of the unit
matrix and everything is trivial.

Conversely, suppose $M_f$ or $M_\nu$ has a mass-independent texture defined by $w$. 
To construct its symmetry operator $G$, we need to find two normalized column vectors
$v'$ and $v''$ which are mutually orthogonal and both orthogonal to $w$. Then the matrix 
\be
G=g_1ww^\dagger+g_2(v'{v'}^\dagger+v''{v''}^\dagger)\ee
obeys $M_fG=GM_f$ for hermitean mass matrices, and $M_\nu G=G^*M_\nu$ for symmetric mass matrices if 
$g_i$ are real. This is so because 
$v'{v'}^\dagger+v''{v''}^\dagger=w'{w'}^\dagger+w''{w''}^\dagger$ if $w'$ and $w''$ are the other two
normalized (pseudo-)eigenvectors of $M$.

In order for $G$ to be unitary, the numbers $g_1$ and $g_2$ must both be a pure phase. For neutrino mass matrix they
also have to be real, so they are $\pm 1$. If by convention we fix $g_1=-1$ and $g_2=1$, then the symmetry
matrix $G$ is unique and $G^2=1$. Of course $-G$ is unitary and it commutes with $M$ as well, so the symmetry group for any Majorana neutrino mass matrix is at least $\G_w=Z_2\times Z_2$, corresponding to the four possible signs of $g_1$ and $g_2$.

 For charged fermions, $g_1$ and $g_2$ could be any phase factor, so the symmetry group is $\G_w=U(1)\times U(1)$. If
we want to limit ourselves to real matrices, then again it comes down to $\G_w=Z_2\times Z_2$.

If the mass matrix has a full texture defined by the vectors $w$ and $w'$, then up to a phase $w''$ is also
known, so that we can take $(v',v'')$ to be $(w',w'')$ for the group $\G_w$, and to be $(w,w'')$ for the group $\G_{w'}$.
The generators of $\G_w$ and $\G_{w'}$ commute with each other, so the symmetry group is then $\G_w\times \G_{w'}$.

If it is known that the mass matrix $M_f$ has the same eigenvalue for its eigenvectors $w'$ and $w''$, then the symmetry group is enlarged
to $\G_w=U(1)\times U(2)$ because the symmetry operator is a more general
\be
G=g_1ww^\dagger+(u_{11}v'{v'}^\dagger+u_{22}v''{v''}^\dagger+u_{12}v'{v''}^\dagger+u_{21}v''{v'}^\dagger),\ee
where $g_1$ is a phase factor and $u_{ij}$ are the elements of any $2\times 2$ unitary matrix $u$.

If the same degeneracy occurs in the Majorana neutrino mass matrix $M_\nu$, then $G$ has to be real to be a symmetry,
so the numbers $g_1$ and $u_{ij}$ have to be taken to be real. The
symmetry group is then $Z_2\times O(2)$, where $O(2)$ is the group of 2-dimensional real orthogonal matrices.
Note that although $SO(2)$ is an abelian group, $O(2)$ itself is non-abelian. 

Since neither the neutrino nor the charged fermion masses are exactly degenerate, these larger symmetries can
only be approximate. Nevertheless, they may be useful in model constructions. An example
of this kind will be discussed below.

Let us illustrate these various possibilities with some examples for the Majorana neutrino mass matrix $M$.

Suppose $M$ has 2-3 texture defined by
$w=(0,1,-1)^T/\r2$, then we can take $v'=(0,1,1)^T/\r2$ and $v''=(1,0,0)^T/\r2$. With $g_1=-1$ and $g_2=1$, 
its unitary symmetry matrix constructed from (6) is
\be
G_{2-3}=\pmatrix{1&0&0\cr 0&0&1\cr 0&1&0\cr}.\ee
This is hardly surprising for a 2-3 texture because it is just the 2-3 permutation matrix. 
Note that if we did not require $G$ to be unitary, then there are many other real
matrices that commute with $M$. For example, taking $g_1=2$ and $g_2=0$, 
\be
G=\pmatrix{0&0&0\cr 0&1&-1\cr 0&-1&1\cr},\ee
is such a matrix.

Next let us assume $M$ to possess the magic texture. Then $w=(1,1,1)^T/\sqrt{3}$ so we can take $v'=(1,-1,0)^T/\r2$ and $v''=(1,1,-2)^T/\rs$.
With $g_1=-1$ and $g_2=1$, its unitary symmetry operator constructed from (6) is
\be
G_{magic}={1\over 3}\pmatrix{1&-2&-2\cr -2&1&-2\cr -2&-2&1\cr}.\ee
If we did not require unitarity, then we may for example choose
$g_1=3$ and $g_2=0$. The resulting democracy matrix \cite{MAGIC1}
\be
G=\pmatrix{1&1&1\cr 1&1&1\cr 1&1&1\cr}\ee
also commutes with $M$, but it is neither unitary nor invertible.

Since $G_{magic}$ is itself 2-3 symmetric, it commutes with $G_{2-3}$. Moreover, $G_{2-3}^2=G_{magic}^2={1}$,
so the symmetry group for the neutrino mass matrix $M_{HPS}$ is $\G_{2-3}\times\G_{magic}$, and 
both $\G_{2-3}$ and $\G_{magic}$ are isomorphic to ${Z}_2\times {Z}_2$, agreeing with the general theory discussed above.

To illustrate the degenerate scenario let us consider the mass matrix \cite{MOHAPATRA}
\be
M=\pmatrix{a&b&b\cr b&a&b\cr b&b&a\cr}.\ee
This mass matrix has a permutation symmetry $S3$ because it is invariant under the exchange of
any two rows and simultaneously the same two columns. It has a non-degenerate eigenvalue $a+2b$ with
eigenvector $v_1=(1,1,1)^T/\rt$, and a doubly degenerate eigenvalue $a-b$ with eigenvectors $v_2=(-1,0,1)^T/\r2$
and $v_3=(1,-2,1)^T/\rs$. Taking $w=v_1$, $v'=v_2$, $v''=v_3$ 
and $g_1=\pm 1$ in (7), the unitary symmetry operator becomes
\be
&&G_{\pm}(u_{11},u_{12},u_{21},u_{22})=\nn\\ &&\nn\\
&&\pmatrix{
\pm{1\over 3}+{1\over 2}u_{11}+{1\over 6}u_{22}-{1\over \rtw}(u_{12}+u_{21})&
\pm{1\over 3}-{1\over 3}u_{22}+{1\over\rt}u_{12}&
\pm{1\over 3}-{1\over 2}u_{11}+{1\over 6}u_{22}-{1\over \rtw}(u_{12}-u_{21})\cr
\pm{1\over 3}-{1\over 3}u_{22}+{1\over\rt}u_{21}&
\pm{1\over 3}+{2\over 3}u_{22}&
\pm{1\over 3}-{1\over 3}u_{22}-{1\over\rt}u_{21}\cr
\pm{1\over 3}-{1\over 2}u_{11}+{1\over 6}u_{22}+{1\over \rtw}(u_{12}-u_{21})&
\pm{1\over 3}-{1\over 3}u_{22}-{1\over\rt}u_{12}&
\pm{1\over 3}+{1\over 2}u_{11}+{1\over 6}u_{22}+{1\over\rtw}(u_{12}+u_{21})\cr}
\nn\\ \ee
where $u_{ij}$ are the matrix elements of a $2\times 2$ real orthogonal matrix.
The non-abelian group $Z_2\times O(2)$ thus generated contains but is much larger than the permutation group $S_3$. 
In particular,
the $S_3$ generators are the identity $G(1,0,0,1)=1$, the two-cycle permutations $G\(\h,\hh,\hh,-\h\)=P_{12},\
G\(\h,-\hh,-\hh,-\h\)=P_{23},\ G\(-1,0,0,1\)=P_{13},$ and the three-cycle permutations $G\(-\h,-\hh,\hh,-\h\)
=P_{123}$ and $G\(-\h,\hh,-\hh,-\h\)=P_{132}$.

\section{Diagonalization Matrix}
Given a textue for a mass matrix $M$, we want to know what restriction
that places on its diagonalization matrix $V$.

The vector $w$ defining a texture is a (pseudo-)eigenvector of $M$.
Let $w'$ and $w''$ be the other two (pseudo-)eigenvectors. Then 
the three columns of $V$ is just the three vectors $w,w',w''$ arranged in some order.

Although we do not know $w'$ and $w''$,
we do know that they can be obtained from any orthonormal pair of basis vectors $v', v''$ 
in the plane orthogonal to $w$ by a unitary rotation.
 We shall let $V_0=(w,v',v'')$ be the unitary matrix whose first, second, and third
columns are given by the vectors $w, v'$, and $v''$, and let
\be
\beta=\pmatrix{\eta&\xi\cr -\xi^*&\eta}\ee
be the unitary matrix that rotates $(v',v'')$ to $(w',w'')$. Namely, $(w',w'')=(v',v'')\beta$. It
is parametrized by a complex number $\xi$ whose norm is not larger than 1, and 
$\eta=\sqrt{1-|\xi|^2}$. 

To write down an explicit mathematical form for $V$, it is convenient to introduce two auxiliary matrices.
Let $P_{jk}=P_{kj}$ be the permutation matrix with 1 in the $(jk), (kj)$ and $(ii)$ entries, and 0 elsewhere
($i\not=j\not=k\not=i$).
Moreover, let 
$P_{ii}$ be the identity matrix. Then $A'=P_{jk}A$
is the matrix obtained by permuting the $j$th and the $k$th rows of $A$, and $A''=AP_{jk}$ is 
the matrix obtained by permuting the
$j$th and the $k$th columns of $A$.

Furthermore, let $B_{jk}(\xi)=B_{kj}(\xi)$ be a block-diagonal unitary matrix with 1 in the $(ii)$ entry, 
0 elsewhere in the $i$th row and 
the $i$th column, and $\beta$ in the $j,k$ rows and columns. Then
the matrix $A'=B_{jk}(\xi)A$ is obtained from $A$ by making a unitary rotation
of its $j$th and $k$th rows, and the matrix $A''=AB_{jk}(\xi)$ 
is obtained from $A$ by making a unitary linear transformation
of its $j$th and $k$th columns.

Putting all these together, we are now ready to write down the general expression for $V$.

First consider the case when the eigenvalue $\l_i$ of $M$ is nondegenerate. Then up to an unimportant phase,
\be
V=V_0P_{1i}B_{jk}(\xi).\ee
$V_0P_{1i}$ is a unitary matrix made up of $w,v',v''$, with $w$ appearing in the $i$th row. $B_{jk}(\xi)$ is
there to rotate $(v',v'')$ into $(w',w'')$.

If $\l_i=\l_j$, then we are also free to have a unitary mixing of the $i$th and the $j$th eigenvectors of $M$,
so the most general form of $V$ is
\be
V=V_0P_{1i}B_{jk}(\xi)B_{ij}(\xi').\ee
Note that the matrix $B_{jk}(\xi)B_{ij}(\xi')$ has a zero in the $(ik)$ position. Conversely, 
any $3\times 3$
unitary matrix $U$ with a zero in the $(ik)$ position can be factorized into the form $B_{jk}(\xi)B_{ij}(\xi')$.

To see that, let $w$ be the $k$th column of $U$, so that $w_i=0$. From the discussion at the end of Sec.~2, we know
that we can find a diagonal matrix $M$ with $M_{jj}=M_{kk}$ to have this texture. Taking $V_0=UP_{1k}$, 
the most
general form for $V$ according to (13) is $V=UB_{ij}(-\xi')B_{jk}(-\xi)$. For later convenience the two parameters
$\xi$ and $\xi'$ in (13) are now called $-\xi'$ and $-\xi$ respectively. The eigenvector $w$ is now placed at
the $k$th column rather than the $i$th column, so that the indices $(i,j,k)$ in (13) become $(k,i,j)$ respectively.
Since $M$ is diagonal, one of these
$V$'s must the the identity. Setting $V$ to be the identity, we see that 
\be
U=B_{jk}(-\xi)^\dagger B_{ij}(-\xi')^\dagger=B_{jk}(\xi)B_{ij}(\xi'),\ee
proving the claim.

Eqs.~(12), (13), and (14) are subject to phase conventions. Since we may alter the phase of each column and each
row at will, we may always pre-multiply each unitary matrix in these equations by a $\Delta(\d_1,\delta_2,\delta_3)
\doteq {\rm diag}(e^{i\d_1},e^{i\d_2},e^{i\d_3})$ to add phases to its rows, 
and post multiply it by a $\Delta(\d_1',\delta_2',\delta_3')$ to add phases to its columns. Of course most of
these phases are not physically meaningful.

In particular, applying (14) to $U=U_{HPS}$ in eq.~(1), whose (13) entry is zero, we get
\be
\Delta(0,0,\pi)U_{HPS}=B_{23}(1/\r2)B_{12}(1/\rt).\ee
We shall return to this factorization later.

Finally, if $\l_i=\l_j=\l_k$, everything is trivial, but we can use the formalism to write a general unitary
matrix $V$ in a factorized form. On the one hand, $M$ is then a multiple of the identity matrix, so $V$
can be any unitary matrix. Moreover, we can choose $V_0P_{1i}={1}$.
On the other hand, since we have three-fold degeneracy in the eigenvalues,
we may mix all three eigenvectors. There is more than one way to mix three eigenvectors, so
$V$ can be written in different ways.

Instead of (13) we now have
\be
V=B_{jk}(\xi)B_{ij}(\xi')B_{ik}(\xi'')=B_{jk}(\xi)B_{ij}(\xi')B_{jk}(\xi''').\ee
The first equality allows a mixing of the $i$th and the $j$th eigenvectors, and then a mixing of the resulting
$i$th and the $k$th eigenvectors. The second expression allows a mixing of the $i$th and the $j$th eigenvectors, and then a mixing of the resulting
$j$th and the $k$th eigenvectors. Again, we may pre- or post- multiply every unitary matrix by a phase matrix $\Delta$.

If $V$ is the mixing matrix, then $V=B_{23}(s_{23})B_{13}(s_{13}e^{-i\d})B_{12}(s_{12})$ is just the Chau-Keung
parametrization \cite{CK}, and $V=B_{23}(-s_2)\Delta(0,0,\delta)B_{12}(-s_1)B_{23}(s_3)$ is just the Kobayashi-Maskawa
parametrization \cite{KM}.

\section{Textures from Mixing Matrices}
Let $V_f$ be the unitary matrix that diagonalizes $M_f=\M_f\M_f^\dagger\ (f=u,d,e)$, so that $V_f^\dagger M_f V_f=\Lambda_f$
is diagonal. Let $V_\nu$ the unitary matrix that diagonalizes
the left-handed Majorana neutrino mass matrix $M_\nu$, so that $V_\nu^TM_\nu V_\nu=\Lambda_\nu$ is diagonal. Then the CKM 
quark mixing matrix is $U_{CKM}=V_u^\dagger V_d$, and  the 
PMNS neutrino
mixing matrix is $U_{PMNS}=V_e^\dagger V_\nu$. 

Mixing matrices can be measured experimentally, but the individual mass
matrices cannot. We want to know what can be said about
the texture of the mass matrices once a mixing matrix is known.

Since the mixing matrix is a product of two diagonalization matrices, there is no way to determine
both of them unless something else is specified. 
We start by assuming both mass matrices to have the same texture, {\it i.e.,} they share some 
symmetry, and ask whether that is
enough to nail them down, and if so, whether this assumption is consistent with experiments.

Let us deal with the neutrino mixing matrix, and assume it to be given
by  $U_{HPS}$ of eq.~(1). If both $M_e$ and $M_\nu$ have
a texture defined by $\tilde w$, then $\tilde w$ is 
an eigenvector of $M_e$ and a pseudo-eigenvector of $M_\nu$. Given any unitary matrix $X$,
$w\doteq X^\dagger \tilde w$ is an eigenvector of $M_e'\doteq X^\dagger M_e X$ and a pseudo-eigenvector
of $M_\nu'\doteq X^TM_\nu X$, hence the transformed mass matrices $M_e'$ and $M_\nu'$
also share a common texture $w$. Their diagonalization matrices are now
$V_e'= X^\dagger V_e$ and $V_\nu'= X^\dagger V_\nu$.

In particular, if we choose $X=V_e$, then $M_e'=\Lambda_e$ is diagonal, 
and the diagonalization matrix for $M_\nu'$ is $V_\nu'=V_e^\dagger V_\nu=U_{PMNS}=U_{HPS}$. 
The vector $w$ defining the common texture comes from one of the columns of $V'_\nu=U_{HPS}$ -- we shall
call the texture $Cp$ if it is taken from column $p$. The matrix $M_e'$ 
shares the same texture and it is diagonal, hence according to
the discussion at the end of Sec.~2, two of its matrix elements must be the same,
$(M'_e)_{jj}=(M'_e)_{kk}$. Moreover, we need to have $w_i=0$ for $i\not=j,k$. Since 
in reality the charged lepton masses are all different,
this mass equality cannot be satisfied, so it is impossible for $M_e'$ and $M'_\nu$, or $M_e$ and $M_\nu$, to
share the same texture. 

However, since the electron and the muon masses are much smaller than the $\tau$ mass, as a first
approximation we might want to regard them to be equal. Even so, the condition $w_i=w_3=0$
cannot be satisfied for any $Cp$ because the third row of $U_{HPS}$ does not contain a vanishing element. 

Similarly, 
if we choose $X=V_\nu$, then $M_\nu'=\Lambda_\nu$ is diagonal, 
and the diagonalization matrix of $M_e'$ is $V_e'=V_\nu^\dagger V_e'=U_{PMNS}^\dagger=U_{HPS}^\dagger$. 
The vector $w$ for the common texture is taken from one of the columns of $V_e'=U_{HPS}^\dagger$, or equivalently,
one of the rows of $U_{HPS}$. We shall label the texture $Rp$ if it is taken from the $p$th row of $U_{HPS}$. 
Since $M_\nu'$ is diagonal and it must
share the same texture $Rp$, the condition
$(M'_\nu)_{ii}=(M'_\nu)_{jj}$ requires the neutrinos to have a two-fold mass degeneracy. This cannot happen because
neither the solar nor the atmospheric gap is zero, so once again we come to the same conclusion that the charged
lepton and the neutrino mass matrices cannot share the same texture. 
 However, since the solar gap is smaller than the atmospheric gap, one might be willing as a first approximation to
assume $(M'_\nu)_{11}=(M'_\nu)_{22}$. In that case we must still require $w_3=0$. The only zero element is the third
column of $U_{HPS}$ is the first one, so we conclude that such an approximate symmetry could be valid,
but only for the texture $R1$. In this way $R1$ distinguishes itself for being
the most symmetrical texture among these six discussed.

A similar discussion can be carried out for the CKM mixing matrix. 
However, since $U_{CKM}$ has no zero element, in order
to obtain an approximate symmetry, not only do we have to assume a two-fold mass degeneracy for the $u$ or the $d$ quarks,
we must also be willing to approximate the small (13) or the (31) elements of the CKM matrix to be zero.

Without the benefit of a common texture, to make any headway
it is necessary to specify how the PMNS matrix is split up into $V_e^\dagger$ and $V_\nu$.
One common practice is to require $V_e=1$, namely, $M_e$ to be diagonal. In that case $V_\nu=U_{HPS}$ and 
$M_\nu$ enjoys the textures $C1,C2$ and $C3$, any two of which implies the third and define a full texture.
The $C2$ texture is just the magic texture and the $C3$ texture is just the 2-3 texture; the $C1$ texture will 
be discussed in the next section. Alternatively, we can require $V_\nu=1$ or $M_\nu$ diagonal. In that case
$V_e=U_{HPS}^\dagger$ and $M_e$ enjoys the $R1,R2,R3$ textures. 

It is also possible to make use of (15) to assign half of $U_{HPS}$ to $V_e^\dagger$ and half to $V_\nu$. 
The textures obtained this way as well as their symmetry operators will be discussed in the next section as well.

These considerations can be generalized to any $U_{PMNS}$. The discussion is similar but the textures
$Rp$ and $Cp$ will be modified. Moreover, we no longer have (15), but using either the 
Kobayashi-Maskawa or the Chau-Keung parametrization of the mixing matrix, we can always factorize the PMNS
matrix into three
factors, two of which can be attributed to $V_e^\dagger$ and one to $V_\nu$, or vice versa.

\section{Neutrino Mixing}

In this section we split the tri-bimaximal mixing matrix $U_{HPS}$ in different ways to discuss the resulting
textures on the leptonic mass matrices. In subsection A, we assume $V_e=1$ and $M_e$ diagonal.
The resulting textures $C1,C2,C3$ 
taken from the three columns of $U_{HPS}$ are discussed one by one, in each case the texture relations, the unitary
symmetry operators, the diagonalization matrix and the Jarlkog invariant are worked out. 
This allows us to parametrize the PMNS matrix if one of
these symmetries is preserved while the other is broken. In subsection B, we assume $V_\nu=1$ and $M_\nu$ diagonal,
and discuss the resulting textures $R1,R2,R3$
taken from the three rows of $U_{HPS}$  in a similar way. As mentioned before, the texture $R1$ is the only
one among these six `diagonal textures' which can be regarded as an approximate symmetry for the diagonal
leptonic mass matrix as well. In subsection C, we consider `factorized textures' by splitting $U_{HPS}$
equally among the two leptonic mass matrices. The two cases considered differ only by a permutation, with
the second case also being the result of a horizontal $Z_2\times S_3$ group with three Higgs doublets in some
mass limits \cite{MONDRAGON}. 

\subsection{$M_e$ diagonal}

\subsubsection{$C1$}
For $C1$ we have
$w=(2,-1,-1)^T/\sqrt{6}$. The resulting texture relations (4) are
\be
2M_{11}+3M_{12}-M_{13}&=&2(M_{22}+M_{23}),\nn\\
2M_{11}-M_{12}+3M_{13}&=&2(M_{23}+M_{33}).\ee
where the symmetry $M_{ij}=M_{ji}$ of the neutrino mass matrix has been used.
 
In particular, if $M$ is diagonal, then (17) requires $M_{11}=M_{22}=M_{33}$, hence the diagonal $M_e$ cannot
possess this texture even if we assume the electron and muon masses to be degenerate. This agrees with the 
general conclusion obtained in the last section.

Setting $V_0=U_{HPS}$ and $i=1$ in eq.~(12), 
the mixing matrix for the $C1$ texture is 
\be
U_{PMNS}=V_\nu=U_{HPS}B_{23}(\xi)=
\pmatrix{2/\rs&\eta/\rt&\xi/\rt\cr -1/\rs&\eta/\rt-\xi^*/\r2&\eta/\r2+\xi/\rt\cr -1/\rs&\eta/\rt+\xi^*/\r2&-\eta/\r2+\xi/\rt\cr}.\ee
The Jarlskog invariant is $J={\rm Im}(U_{11}U_{22}U_{12}^*U_{21}^*)=-(\rs/18)\eta\ {\rm Im}(\xi)$, where $U$ stands
for $U_{PMNS}$.

The unitary symmetry operator $G$ for this texture can be obtained from (6), by setting $g_1=-1$ and $g_2=1$
as per our convention discussed in Sec.~3. The vectors $v'$ and $v''$ can be taken from the second and third
columns of $U_{HPS}$. The result is 
\be
G_{C1}={1\over 3}\pmatrix{-1&2&2\cr 2&2&-1\cr 2&-1&2\cr}.\ee

\subsubsection{$C2$}
For $C2$ we have $w=(1,1,1)^T/\rt$. This is just the magic texture whose texture relations
and symmetry operator $G_{C2}=G_{magic}$ have been discussed in Sec.~3. Again, the row sums
of the diagonal $M_e$ are all equal only when all the charged lepton masses are equal, so
$M_e$ cannot possess this texture even in the approximation $m_e=m_\mu$, as concluded in the last section.

With $V_0P_{1i}=U_{HPS}$, 
the mixing matrix for the $C2$ texture is 
\be
U=U_{HPS}B_{13}(\xi)=
\pmatrix{\sqrt{2/3}\eta&1/\sqrt{3}&\sqrt{2/3}\xi\cr
-\eta/\sqrt{6}-\xi^*/\r2&1/\sqrt{3}&\eta/\sqrt{2}-\xi/\rs\cr
-\eta/\sqrt{6}+\xi^*/\r2&1/\sqrt{3}&-\eta/\sqrt{2}-\xi/\rs\cr}.\ee
This parametrization was first obtained in
\cite{BHS} with $\sqrt{2/3}\xi=u$.
The Jarlskog invariant is $-\sqrt{2/3}\ {\rm Im}(\xi)$.

\subsubsection{$C3$}
For $C3$ we have $w=(0,-1,1)^T/\sqrt{2}$. This is just the 2-3 texture whose texture relations
and symmetry operator $G_{C3}=G_{2-3}$ have been discussed in Sec.~3. Since $m_\mu\not=m_\tau$, the diagonal
$M_e$ cannot possess this texture.

With $V_0P_{1i}=U_{HPS}$, 
the mixing matrix for the $C3$ texture is 
\be
U=U_{HPF}B_{12}(\xi)=\pmatrix{2\eta/\rs -\xi^*/\rt&2\xi/\rs +\eta/\rt&0\cr
-\eta/\rs-\xi^*/\rt&-\xi/\rs+\eta/\rt&1/\r2\cr -\eta/\rs-\xi^*/\rt&-\xi/\rs+\eta/\rt&-1/\r2\cr}.\ee
The Jarlskog invariant
\be
J={\rm Im}(U_{22}U_{23}^*U_{32}^*U_{33})\ee
is zero in this case, so we might as well take $\xi$ to be real. In that case, with $s$ real and $c=\sqrt{1-s^2}$
real, we can rewrite $U$ in a more familiar form
\be
U=\pmatrix{c&s&0\cr -s/\r2&c/\r2&1/\r2\cr -s/\r2&c/\r2&-1/\r2\cr}.\ee
Thus in the presence of a 2-3 texture, the reactor angle is zero, the atmospheric mixing is maximal, but
the solar angle remains to be free.

\subsection{$M_\nu$ diagonal}

\subsubsection{$R1$}

For $R1$, take $w=(\r2,1,0)^T/\rt$. Then (4) becomes
\be
M_{12}&=&\r2(M_{11}-M_{22}),\nn\\
M_{23}&=&-\r2 M_{13}.\ee
A diagonal matrix may satisfy these relations provided $M_{11}=M_{22}$, hence if we are allowed to ignore the solar gap,
then the diagonal $M_\nu$ may possess this texture approximately, as concluded in the last section.
This is the only texture among all the $Ri$'s and $Ci$'s that has this property of being approximately common to
both the leptonic mass matrices.

Setting $V_0P_{1i}=U_{HPS}^T$, we get from (12) that 
\be
U_{PMNS}=V_e^\dagger=B_{23}(\xi)^\dagger U_{HPS}=
\pmatrix{\r2/\rt&1/\rt&0\cr (\xi-\eta)/\rs&(\eta-\xi)/\rt&(\eta+\xi)/\r2\cr -(\xi^*+\eta)/\rs&(\xi^*+\eta)/\rt&(\xi^*-\eta)/\r2\cr}.
\ee
Since one of the matrix elements is zero, the Jarlskog invariant $J=0$, hence we might as well let $\xi$ to be real.
According to the breaking pattern of (25), the solar and the reactor angles maintain their HPS values, but the
atmospheric angle can change.

The unitary symmetry operator is
\be
G_{R1}={1\over 3}\pmatrix{-1&-2\r2&0\cr -2\r2&1&0\cr 0&0&1\cr}.\ee
\subsubsection{$R2$}
For $R2$, take $w=(-1,\r2,\rt)^T/\rs$. Then (4) becomes
\be
\r2(M_{11}-M_{22})-M_{12}-\rs M_{13}-\rt M_{23}&=&0,\nn\\
\rt(M_{11}-M_{33})-\rs M_{12}-2M_{13}-\r2 M_{23}&=&0.\ee
For the diagonal $M_\nu$ to share this texture, both the solar and the atmospheric gaps must be put to zero.

With $V_0P_{1i}=U_{HPS}^T$, we get from (12) that 
\be
U_{PMNS}=V_e^\dagger=B_{13}(\xi)^\dagger U_{HPS}=
\pmatrix{(2\eta+\xi)/\rs&(\eta-\xi)/\rt&\xi/\r2\cr
-1/\rs&1/\rt&1/\r2\cr  (-\eta+2\xi^*)/\rs& (\eta+\xi^*)/\rt& -\eta/\r2\cr}.
\ee
The Jarlskog invariant is $-\eta{\rm Im}(\xi)/6$, and the unitary symmetry operator is
\be
G_{R2}={1\over 3}\pmatrix{2&\sqrt{2}&\rt\cr  \sqrt{2}&1&-\rs\cr \sqrt{3}&-\rs&0\cr}.\ee

\subsubsection{$R3$}
For $R3$, 
take $w=(-1,\r2,-\rt)^T/\rs$. Then (4) becomes
\be
\r2(M_{11}-M_{22})-M_{12}+\rs M_{13}+\rt M_{23}&=&0,\nn\\
\rt(M_{11}-M_{33})-\rs M_{12}+2M_{13}+\r2 M_{23}&=&0.\ee
Again, for the diagonal $M_\nu$ to share this texture, both the solar and the atmospheric gaps must be put to zero.

Setting $V_0P_{1i}=U_{HPS}^T$, we get from (12) that 
\be
U_{PMNS}=V_e^\dagger=B_{12}(\xi)^\dagger U_{HPS}=
\pmatrix{(2\eta+\xi)/\rs&(\eta-\xi)/\rt&-\xi/\r2\cr
  (-\eta+2\xi^*)/\rs& (\eta+\xi^*)/\rt& \eta/\r2\cr -1/\rs&1/\rt&-1/\r2\cr}.
\ee
The Jarlskog invariant is $\eta{\rm Im}(\xi)/6$ and the unitary symmetry operator is
\be
G_{R3}={1\over 3}\pmatrix{2&\sqrt{2}&-\rt\cr  \sqrt{2}&1&\rs\cr -\sqrt{3}&\rs&0\cr}.\ee

\subsection{Factorized textures}
Since $\Delta(0,0,\pi)U_{HPS}$ has the factorized form (15), we may let
\be
V_e^\dagger=B_{23}(1/\r2)Y^\dagger=\pmatrix{1&0&0\cr 0&1/\r2&1/\r2\cr 0&-1/\r2&1/\r2\cr}Y^\dagger,\ee
 and
\be V_\nu=YB_{12}(1/\rt)=Y\pmatrix{\r2/\rt&1/\rt&0\cr -1/\rt&\r2/\rt&0\cr 0&0&1\cr},\ee
where $Y$ is some unitary matrix. We have choosen a different phase convention 
[$U_{PMNS}=\Delta(0,0,\pi)U_{HPS}$] in this subsection for later convenience.

Without specifying $Y$, this is completely general. We shall discuss
in this subsection two examples in which $Y$ is chosen to maintain the number of zeros in both $V_e$ and $V_\nu$.

\subsubsection{$Y={\bf 1}$}
From (4), we obtain the full textures of $M_e$ and $M_\nu$ to be
\be
(M_e)_{12}=(M_e)_{13}=(M_e)_{22}-(M_e)_{33}=0,\ee
and
\be
(M_\nu)_{13}=(M_\nu)_{23}=(M_\nu)_{11}-(M_\nu)_{22}+{1\over\r2}(M_\nu)_{12}=0.\ee
Explicity, these mass matrices are of the form
\be
M_e&=&\pmatrix{a&0&0\cr 0&b&c\cr 0&c&b\cr},\\ &&\nn\\
M_\nu&=&\pmatrix{d&\r2(e-d)&0\cr \r2(e-d)&e&0\cr 0&0&f\cr}.\ee
Using (33) and (34), we can relate these parameters to the fermion masses. The results are
\be
a=m_e^2,\quad b=(m_\mu^2+m_\tau^2)/2, \quad c=(-m_\mu^2+m_\tau^2)/2,\ee
and
\be
d=(2m_{\nu_1}+m_{\nu_2})/3,\quad e=(2m_{\nu_2}+m_{\nu_1})/3,\quad f=m_{\nu_3}.\ee

Since both $M_e$ and $M_\nu$ have a full texture, each has two symmetry operators. 
The symmetry operators for $M_e$ computed from (6) and (37) are
\be
G_e=\pmatrix{-1&0&0\cr 0&1&0\cr 0&0&1\cr},\quad G_e\! '=\pmatrix{1&0&0\cr 0&0&1\cr 0&1&0\cr},\ee
and the symmetry opreators for $M_\nu$ computed from (6) and (38) are
\be
G_\nu=\pmatrix{1&0&0\cr 0&1&0\cr 0&0&-1\cr},\quad G_\nu\! '=\pmatrix{1/3&-2\r2/3&0\cr -2\r2/3&-1/3&0\cr 0&0&1\cr}.\ee

\subsubsection{$Y=P_{23}P_{12}$}
This is similar to the previous case, except that a permutation (132) is applied to the rows of $V_e$ and $V_\mu$.
Hence
\be
V_e=\pmatrix{ 0&1/\r2&1/\r2\cr 0&-1/\r2&1/\r2\cr 1&0&0\cr},\ee
 and
\be V_\nu=\pmatrix{ -1/\rt&\r2/\rt&0\cr 0&0&1\cr \r2/\rt&1/\rt&0\cr}.\ee
The full textures of $M_e$ and $M_\nu$ are then obtained from (37) and (38) by making a (132) permutation to
their rows and columns:
\be
M_e&=&\pmatrix{b&c&0\cr c&b&0\cr 0&0&a\cr},\\  &&\nn\\
M_\nu&=&\pmatrix{e&0&\r2(e-d)\cr 0&f&0\cr \r2(e-d)&0&d\cr}.\ee
The symmetry operators in (41) and (42) will also be similarly permuted.
 
Using (43)-(46), the relation between the fermion masses and the mass matrix parameters is once again
given by (39) and (40). Hence
\be
M_e=\pmatrix{(m_\mu^2+m_\tau^2)/2&(m_\tau^2-m_\mu^2)/2&0\cr (m_\tau^2-m_\mu^2)/2&(m_\mu^2+m_\tau^2)/2&0\cr 0&0&m_e^2\cr},\ee
and
\be
M_\nu=\pmatrix{(2m_{\nu_2}+m_{\nu_1})/3&0&\r2(m_{\nu_2}-m_{\nu_1})/3\cr
0&m_{\nu_3}&0\cr \r2(m_{\nu_2}-m_{\nu_1})/3&0&(2m_{\nu_1}+m_{\nu_2})/3\cr}.\ee

This coincides with the result of a ${Z_2}\times S_3$ horizontal symmetry \cite{MONDRAGON} in the limit $m_e=0$ and 
$m_{\nu_3}=(2m_{\nu_2}+m_{\nu_1})/3$.

\section{Conclusion}
We have investigated the construction and the general property of mass-independent textures
(Sec.~2), and showed that they can be interpreted as
 symmetries of mass matrices (Sec.~3). An explicit recipe is given to construct the unitary symmetry operators 
and symmetry groups (Sec.~3),
 together with several illustrative examples (Sec.~6). We found the symmetry group corresponding
to any simple texture for a charged fermion mass matrix to be $U(1)\times U(1)$, and that for a Majorana neutrino
mass matrix to be $Z_2\times Z_2$. If the mass matrix has a degenerate eigenvalue, then its symmetry group
is $U(1)\times U(2)$ and $Z_2\times O(2)$ respectively.
Whatever a texture is, we found that
both members of an isodoublet cannot simultaneously possess the same texture, though they may do
so approximately for the texture $R1$ of the neutrino tri-bimaximal mixing matrix (Sec.~5). 
Various textures arriving from the tri-bimaximal neutrino mixing matrix were considered, together with
the symmetries they satified. This includes the six `diagonal textures' $C1, C2, C3$ and $R1,R2,R3$, as well as two
`factorized textures', one of which coincides with the result of a minimal $Z_2\times S_3$ horizontal symmetry
in some mass limit (Sec.~6).
We have also investigated possible patterns of symmetry breaking from the tri-bimaximal neutrino mixing,
assuming one of the horizonal symmetries to remain intact while the others are broken (Sec.~6). 

This research is supported by the Natural Sciences and Engineering Research Council of Canada.


\end{document}